\documentclass[twocolumn,pre,aps,superscriptaddress,showpacs,epsfig,dvips,amsmath,amssymb]{revtex4}
\usepackage{eufrak}
\usepackage[dvips]{graphicx}
\usepackage{dcolumn}
\usepackage{epsfig}
\usepackage{bm}
\usepackage{multirow}
\usepackage{color}

\begin{document}

\title{Vacancy diffusion in the triangular lattice dimer model}
\date{\today}
\author{Monwhea Jeng}
\affiliation{Department of Physics, Syracuse University, Syracuse, NY 13244}
\author{Mark J. Bowick}
\affiliation{Department of Physics, Syracuse University, Syracuse, NY 13244}
\author{Werner Krauth}
\affiliation{CNRS-Laboratoire de Physique Statistique, Ecole Normale Sup\'erieure,24 rue Lhomond, 75231 Paris Cedex 05, France}
\author{Jennifer Schwarz}
\affiliation{Department of Physics, Syracuse University, Syracuse, NY 13244}
\author{Xiangjun Xing}
\affiliation{Department of Physics, Syracuse University, Syracuse, NY 13244}
\email{xxing@physics.syr.edu}

\date{\today} 

\begin{abstract} 
We study vacancy diffusion on the classical triangular lattice 
dimer model, subject to the kinetic constraint that dimers 
can only translate, but not rotate. A single vacancy, i.e. 
a monomer, in an otherwise fully packed lattice, is always 
localized in a tree-like structure. The distribution of 
tree sizes is asymptotically exponential and has an average 
of $8.16\pm 0.01$ sites. A connected pair of monomers has a finite 
probability of being delocalized. When delocalized, the 
diffusion of monomers is anomalous: 
$\langle \vec{x}^2 \rangle \propto t^{\beta}$, with
$\beta=0.46\pm 0.06$. We also find that the same exponent $\beta$
governs diffusion of clusters of three or four monomers, 
as well as the diffusion of dimers at finite but low monomer densities. 
We argue that coordinated motion of monomer pairs is the basic mechanism
allowing large-scale transport at low monomer densities. We further 
identify a ``swap-tunneling'' mechanism for diffusion 
of monomer pairs, where a subtle interplay between swap 
moves (translations of dimers transverse to their axes) 
and glide moves (translations of dimers parallel to their 
axes) plays an essential role.
\end{abstract} 

\pacs{05.50.+q,
68.55.Ln,
45.70.-n 
}
\maketitle


\section{Introduction}
\label{Sec:intro}
The statistical mechanics of the lattice dimer model has a long
and venerable history~\cite{dimer:Lieb,dimer:Kenyon}. It is
one of the earliest prototypical lattice models where hard
constraints play an essential role and that has interesting and
deep connections with the Ising model and various kinds of
lattice gauge theories~\cite{PhysRevB.65.024504}. It has
been extensively studied in the setting of random sequential
adsorption processes (RSA), both reversible and irreversible~\cite{review-RSA}.
More recently, interests in dimer models have been further boosted 
by their relevance to the resonance valence bond (RVB) theory of 
high $T_c$ superconductivity~\cite{RVB:Anderson,dimer:RK}.

The equilibrium physics of lattice dimer models is already 
well understood. The partition function of fully packed 
dimers on any planar lattice can be exactly computed using 
the Pfaffian method, following a theorem of Kasteleyn~\cite{dimer:Kasteleyn-1,dimer:Kasteleyn-2,dimer:Fisher-1,dimer:Fisher-2,dimer:Fisher-3}. 
The case with a nonzero monomer fraction proves to be more 
difficult and interesting. Both analytic techniques and 
numerical simulations have been used to attack this problem 
\cite{dimer-triangular-Fendley,pocket-algor:Krauth}. Previous studies of 
two-dimensional equilibrium dimer models 
seem to suggest two universality classes~\cite{pocket-algor:Krauth}: 
For bipartite lattices, monomers on different sublattices 
behave as positive or negative charges, interacting with 
a logarithmic Coulomb potential of entropic origin. The 
physics of a finite monomer density system is, therefore, well described 
by the Debye-Huckle theory of a 2D plasma. For non-bipartite 
lattices, however, the monomers behave as a weakly 
interacting gas with extremely short-range correlations. 
This distinction between bipartite and non-bipartite lattices 
seems to persist even in three dimensions~\cite{dimer-Krauth-3d}. 

Appropriately defined dynamics of dimers may describe the structural 
rearrangement in dense anisotropic granular or glassy
systems.  Furthermore, the fact that the equilibrium physics of the dimer model 
is well understood makes it particularly convenient for a dynamic study. 
The similarities between glasses and dense granular systems have long been 
recognized and explored.  Furthermore, recent theoretical studies on the glassy 
dynamics of lattice models with point-like particles, such as the 
Kob--Anderson model~\cite{Kob.Andersen.Model,TBF-KA-1,TBF-KA-2} 
and other more exotic models~\cite{TBF-JP,TBF-JS-comment,TBF-reply}, 
have revealed deep connections between kinetic constraints and glassy
dynamics, as well as new mechanisms of ergodicity breaking in lattice 
systems~\footnote{This ergodicity-breaking transition is believed to become 
a crossover in a continuous space model.}.    It is therefore interesting to explore 
how kinetic constraints affect the diffusion of dimers as well as vacancies in the 
dimer model.   Two studies of single-monomer diffusion in an otherwise fully 
occupied square lattice have been published 
recently~\cite{Square.Dimer.Monomer,poghosyan:041130}. 
Here we focus on the two-dimensional triangular lattice, and study diffusion 
of both single-monomer and monomer clusters.

\subsection{Model and Summary of Results}

Our main goal is to characterize, both qualitatively and 
quantitatively, the diffusion of vacancies, i.e. monomers, 
in a densely packed lattice dimer model, when the dynamics 
are subject to hardcore repulsion, as well as to various
kinetic constraints. We only allow single dimer moves that 
do not cause double occupation at any site at any time. This naturally
excludes two-dimer dynamics such as those considered in the
context of quantum dimer models. Therefore no dimer can
move in a fully packed triangular lattice; i.e. the system
is completely jammed.

\begin{widetext}
\begin{center}
\begin{figure}[tb]
\epsfig{figure=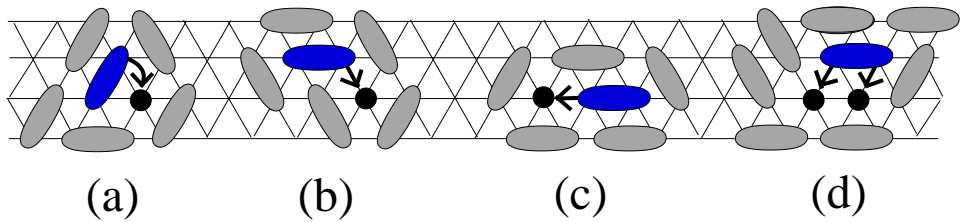,width=5.0in}
\caption{Types of dimer moves. (a) and (b) are rotations, 
and are not allowed in our model.
(c) shows a glide move, and (d) shows a swap move.}
\label{fig:allowed.moves}
\end{figure}
\end{center}
\end{widetext}

A dimer can only move when there are one or more vacancies,
i.e. monomers, in its immediate neighborhood. When a dimer
has only one nearest monomer, there are two kinds of
moves that we could allow: moves in which the dimer changes 
its orientation as well as its position, which we call a rotation, or
moves in which it simply translates along its axis, which we call a 
glide (see Fig.~\ref{fig:allowed.moves}a-c). If both rotations and 
glides of dimers are allowed, a monomer can always move in 
any of the six possible directions. Therefore, a single monomer 
in an otherwise fully occupied triangular lattice simply performs 
a random walk.  At finite monomer density, all monomers
simply behave as weakly interacting gas molecules. The
dimer diffusion constant scales linearly with the monomer
density $\rho_m$. Such a scenario is clearly uninteresting.

\begin{figure}[tb]
\epsfig{figure=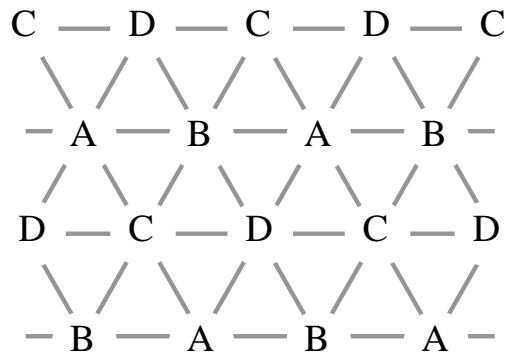,width=2.8in}
\caption{The four sub-lattices in the dimer model.}
\label{fig:sublattice}
\end{figure}

We shall therefore forbid rotations of dimers from now 
on~\footnote{Forbiddance of rotational degrees of freedom 
may also be relevant to the glassy physics of anisotropic 
liquids.  It has been widely recognized~\cite{Fujara-92,PhysRevLett.90.015901,kivelson:4464,PhysRevLett.79.103}, for example, that  
rotational relaxation becomes much slower than translational relaxation 
near the glass transition.  It is therefore possible 
that suppression of rotational motion results in glassy dynamics. }. 
Given an isolated monomer, if there is a nearest dimer with its axis 
pointing towards the monomer, as shown in 
Fig.~\ref{fig:allowed.moves}c, the dimer can glide into the
monomer site.  In this case, the monomer moves along one of
the three crystal axes by two lattice steps. Therefore,
with glide moves a 
monomer can only move on one of the four sub-lattices of the
triangular lattice, shown in Fig.~\ref{fig:sublattice}.
Furthermore, we shall prove in Sec.~\ref{Sec:single-monomer} 
that a monomer can never return to its initial lattice site from a
different direction than which it left it.  Therefore all lattice sites 
reachable by an isolated monomer form a tree-like structure, 
which we shall call a {\em monomer tree}.
Both back-of-the-envelope calculations and numerical simulations
indicate that monomer tree sizes are always finite. The distribution 
of the monomer tree size is asymptotically exponential, as illustrated 
in Fig.~\ref{fig:treesizes}, with an average of $8.16\pm0.01$ sites.
We thus expect that single monomer moves do not contribute to the
large-scale transport of dimers at high packing density.
Consequently, some other collective mechanism is needed for
the diffusion of dimers over large length scales.

Now let us consider two monomers which are nearest neighbors
to each other \footnote{If two monomers can not become 
nearest neighbors to each other by glide moves, they simply 
behave as two independent and isolated monomers, 
which do not contribute to the large-scale diffusion of dimers.}. 
We allow nearby dimers to translate transverse to their axes and 
occupy the lattice sites of two monomers, as illustrated in 
Fig.~\ref{fig:allowed.moves}d.  Such a dimer move shall be 
called a {\em swap}.  Swap moves provide a mechanism for 
changing the monomer tree structures, which is essential for 
large-scale transport in the triangular lattice dimer model in 
the high packing density regime. 
Nevertheless, glide moves separate monomer pairs that are
nearest neighbors to each other, and make swap moves unavailable. 
Once separated, two monomers can form a nearest neighbor pair 
again only at one or more particular pairs of sites. These reconnection 
events are clearly suppressed by entropic barriers.
We therefore have the following qualitative picture for the diffusion of
a two-monomer cluster: each monomer may diffuse on its
individual monomer tree, via glide of dimers. This move
is entropically preferred, but does not contribute to the
large-scale transport of monomers/dimers. Only occasionally, 
the two monomers meet neighboring sites,
whereupon they may be able to
travel together by a swap move. Each monomer
then discovers a new monomer tree on which it can diffuse.

\begin{figure}[tb]
\epsfig{figure=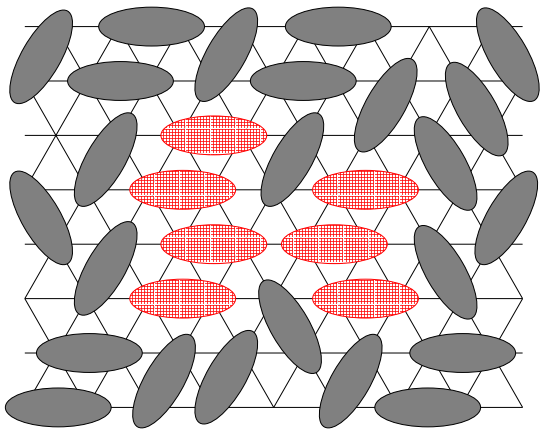,width=2.8in}
\caption{(Color online) A swap cluster. If a two-monomer cluster is created 
by removing one dimer inside this swap cluster, it can reach 
all sites in the swap cluster by swap or double-glide moves 
of dimers.} 
\label{fig:swap.cluster.picture}
\end{figure}

\begin{figure}[tb]
\epsfig{figure=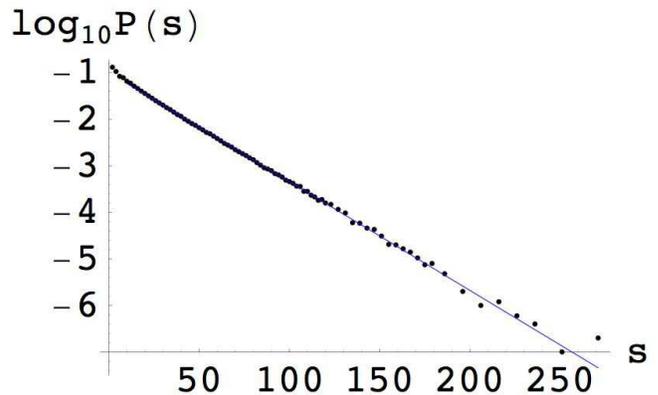,width=3.4in}
\caption{(Color online) Probability distribution of swap cluster sizes.
The x-axis measures the number of sites in the swap cluster. }
\label{fig:swap.distribution}
\end{figure}

One might suspect that two-monomer clusters can diffuse 
faster if we forbid altogether glide moves of dimers.  This turns out, 
however, not to be true.  Let us first define a {\em swap cluster} in
a fully packed lattice as the maximal subset of dimers that have 
the same orientation, such that if any of these dimers are removed, 
the resulting connected monomer pair can visit the place of any
other dimer in the same cluster by swap moves or double-glide moves, 
i.e. swap moves or two consecutive glide moves along the same 
direction.   An example of a swap cluster is shown in 
Fig.~\ref{fig:swap.cluster.picture}.  We have checked numerically that 
in the equilibrium ensemble, all swap clusters are finite.  
As shown in Fig.~\ref{fig:swap.distribution}, the distribution of swap 
cluster sizes is exponential, on average covering $17.89\pm 0.02$ sites. 
This behavior is of course in qualitative agreement with 
the extremely short-range correlations exhibited by the equilibrium 
ensemble of the triangular lattice dimer model. 

\begin{figure}[tb]
\epsfig{figure=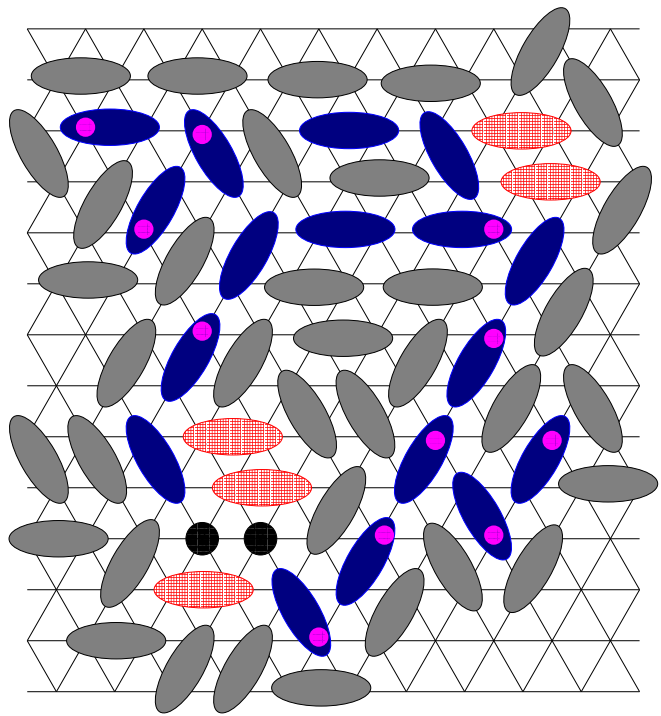,width=3.4in}
\caption{(Color online) A configuration in which two monomer trees touch 
at a position outside their original swap cluster.
The monomer pair starting from the bottom left swap cluster
can ``tunnel'' to the other swap cluster on the top right
(to the sites enclosed by the brown circle).
The initial and final swap clusters are indicated by the
dimers filled
with (red) hatched lines. The dark (blue) dimers and the 
small, light (pink) dots
indicate the monomer trees of the individual monomers (see
section~\ref{Sec:single-monomer}).}
\label{fig:swap.picture.treestouch}
\end{figure}

\begin{figure}[tb]
\epsfig{figure=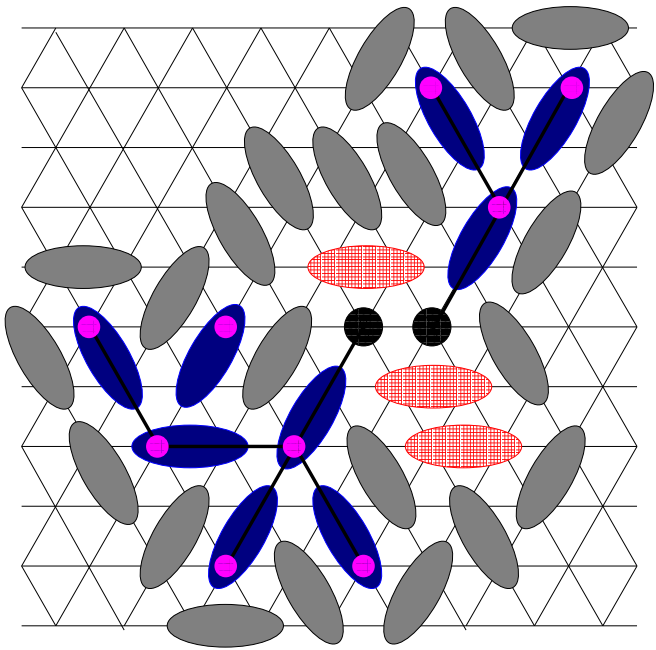,width=3.4in}
\caption{(Color online) A configuration in which two monomer trees touch 
only at the original positions (two black dots) of the monomers.
For this monomer pair to move large distances, the monomers 
have to meet each other at the two black dots (a reconnection 
event) and then swap to another location in the swap cluster.
The swap cluster is indicated by the dimers filled 
with (red) hatched lines. The dark (blue) dimers and the 
small, light (pink) dots
indicate the monomer trees of the individual monomers (see
section~\ref{Sec:single-monomer}).}
\label{fig:swap.picture.treesapart}
\end{figure}

Now if we create a two-monomer cluster by removing a dimer 
inside a swap cluster, and only allow swap and double-glide
moves, by definition the monomer pair can only visit all the sites 
of the swap cluster--all sites that are reachable by swap 
or double-glide move belong to the same swap cluster. 
That is, the monomer pair is {\em localized} if 
dimer glides are not allowed. Without assistance from other
monomers, a monomer pair can escape from a swap cluster only
by one mechanism: two monomers may ``tunnel'' through their
individual monomer trees by glide moves and rejoin each other inside 
some other swap cluster. A configuration in which such tunneling is 
possible is shown in Fig.~\ref{fig:swap.picture.treestouch}.
Another configuration in which such tunneling is not possible 
(without first carrying out a swap move) is shown in 
Fig.~\ref{fig:swap.picture.treesapart}.
We therefore deduce that {\em both glide and swap moves 
are essential for large-scale diffusion of monomers.} 

The probability that a monomer pair can escape from a swap
cluster depends on the details of dimer packing around the
monomer pair. It is not a priori clear whether a monomer pair
can diffuse around the whole system by this mechanism. We
have run extensive simulations of
the diffusion of monomers, and found
numerical evidences which show that, when randomly
prepared, a finite fraction (about 20\%) of monomer pairs are localized,
while the remaining fraction can diffuse to infinity.
We have also simulated monomer clusters consisting of three
or four monomers, and found they are almost always delocalized.
Furthermore, we found that in all these cases, the monomer
diffusion is anomalous, with an exponent of $0.46\pm0.05$. 
We are, however, not able to find a quantitative
understanding of this diffusion law.  Finally, we have also simulated
diffusion of dimers at finite but low monomer densities and discovered 
also the same anomalous diffusion exponent $\approx 0.47$. 
This anomalous diffusion of dimers can be understood in terms 
of diffusion of monomer pairs. 

The remainder of this paper is organized as follows. In
Sec.~\ref{Sec:method} we discuss some details of the model and
the numerical method we used in our simulations. In Sec.
\ref{Sec:single-monomer} we study the diffusion (localization)
of a single monomer in an otherwise fully packed lattice. In
Sec.~\ref{Sec:localization} we present the results on
the localization and delocalization
of clusters of two, three, or more
monomers.
In Sec.~\ref{Sec:diffusion}
we analyze the anomalously slow diffusion of monomers,
as well as the statistics of monomer pair reconnection events. 
In Sec.~\ref{Sec:dimer-diffusion}
we look at the diffusion of dimers in states with finite
monomer density.


\section{Simulation methods}
\label{Sec:method}

To prepare the appropriate initial random state of an $L\times L$ triangular 
lattice packed with dimers, with periodic boundary conditions, we use 
the pocket algorithm~\cite{pocket-algor:Krauth,algorithm:Krauth,SM:Krauth}.
If we want a configuration 
with an even number of monomers, we start with a fully packed 
and fully ordered state (all dimers in the same direction) 
on an even-by-even lattice; for a configuration with 
an odd number of monomers, we start with an odd-by-odd lattice 
that has only one monomer, and that is as nearly fully ordered 
as possible. We then randomize the state with the 
pocket algorithm~\cite{pocket-algor:Krauth,algorithm:Krauth,SM:Krauth}. 
This algorithm is ergodic, and satisfies detailed balance with respect 
to the trivially flat measure in configuration space.  Each 
iteration of the algorithm rearranges a large number of 
dimers, so that the system quickly reaches a random state. 
A large number of pivots ($5L^2$) are carried out to ensure 
reaching equilibrium. After that, a smaller number of pivots 
($10 L$) are successively applied to produce other independent 
random states. It is already known that these random states 
have only short-range correlations in the dimer 
orientations~\cite{dimer-triangular-Fendley,pocket-algor:Krauth}. 

Given a fully packed state (on an even-by-even lattice) or 
a one-monomer state (on an odd-by-odd lattice), we then
generate states with more monomers by removing dimers. 
When generating states with three or four monomers, we remove
dimers adjacent to the already-existing monomers, to create
larger monomer clusters. 

As stated earlier, in this model dimers can make both
glide and swap moves (Fig~\ref{fig:allowed.moves}c and d).
Once we have an initial state, we carry out these moves. 
We set the time scale such that, on average, over every unit
of time, every dimer attempts one move, choosing at random one
of the six possible directions available to it. The
attempted move is carried out if and only if it satisfies the
hard-core constraint; for an glide move, the site the dimer
is moving into needs to be vacant, while for a swap move,
both of the sites need to be vacant. 

Since we are looking at configurations with low numbers of monomers,
generating trial moves by picking among the dimers at random
is inefficient. We therefore use the following equivalent, but more
efficient algorithm: if we have $N$ monomers, then every 
step we advance the time by $1/N$, 
pick a random monomer, and a random site adjacent to that
monomer. If the adjacent site is occupied by a dimer that can
move into the monomer with an glide move, we do so. 
If the adjacent site is occupied by a dimer that can move into 
the monomer with a swap move, we do so with probability 1/2. 
This generates moves with the same probabilities as if we instead 
chose random dimers.


\section{Dynamics of Single Monomer}
\label{Sec:single-monomer}

We first consider configurations with only a single monomer.
With only one monomer in the system, swap moves can never occur; 
only glide moves are allowed. 
With every glide move, the monomer moves two spaces, so
the monomer is confined
to one of four sub-lattices---see Fig.~\ref{fig:sublattice}.
This in turn means that a dimer cannot make two glide
moves in the same direction---that is, make a glide move
in a direction and then, after some moves of 
other dimers, make a second glide move in the same
direction---because it would then be moving into vacancies
of different sub-lattices in the two steps, contradicting
the result that the monomer remains on the same
sub-lattice. This means that with only a single monomer, a
dimer is either confined to moving back and forth between
two positions, or not moving at all.

This in turn means that for configurations with a single monomer,
the latter can never move in a loop (come back to its
original lattice site from a different direction than it
left that site). Since dimers can only carry out ``back and
forth'' moves, after the monomer leaves its original site
by moving a certain dimer, it can only return to its
original site from the direction that it left it, by moving the 
same dimer back to its original position.

\begin{figure}[tb]
\epsfig{figure=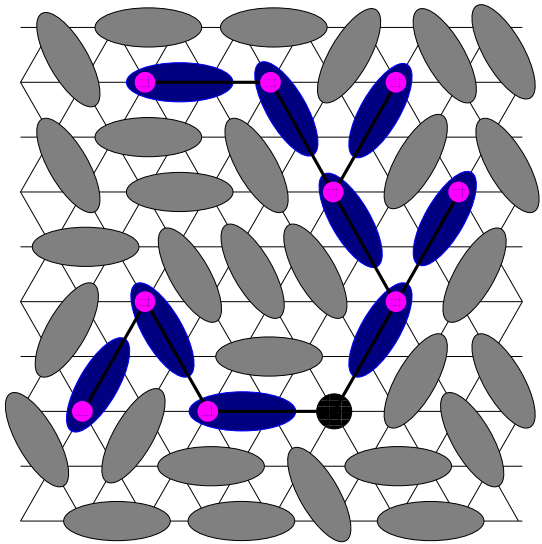,width=2.8in}
\caption{(Color online) A monomer tree. The dark (blue) dimers 
can eventually move, 
and the small, light (pink) dots show the sites that the monomer can reach.
 }
\label{fig:monomer.tree}
\end{figure}

This means that for a state with a single monomer, it is
easy to quickly figure out which sites the monomer can
reach, without explicitly carrying out the dimer moves. 
We can first see which sites a monomer can reach with a
single glide move by inspecting the orientation of dimers
on adjacent sites. If a dimer on a neighboring site points
towards the monomer, then it can make a glide move into the
monomer. 
Once the monomer makes a single move, it can either move
back in the direction that it came from, or make a new move. We
can determine what new moves are possible by the same
process as before, and thus construct the set of sites that 
the monomer can reach by repeated glide moves.
Since the monomer can never move in a loop, the set of sites
that the monomer can reach forms a {\em static} tree, such as 
the one shown in Fig.~\ref{fig:monomer.tree}. 
An isolated monomer thus performs a random walk on its tree. 

\begin{figure}[tb]
\epsfig{figure=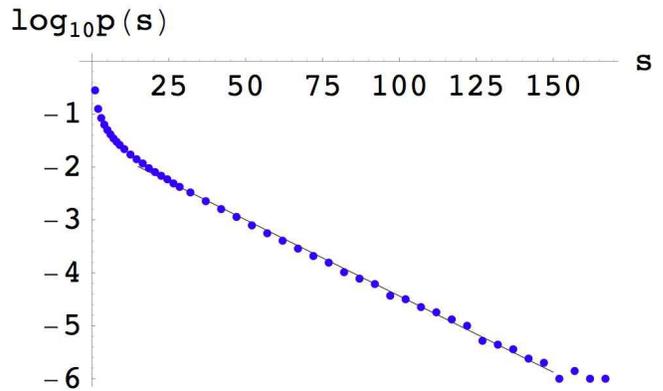,width=3.4in}
\caption{Distribution of monomer tree sizes (the number of sites $s$ 
a single monomer can reach) in a $101\times 101$ lattice with a
single vacancy. }
\label{fig:treesizes}
\end{figure}

Our numerical simulations find that large monomer trees are 
exponentially suppressed. Fig.~\ref{fig:treesizes} shows the
distribution of monomer trees sizes for $10^5$ configurations 
on a $101\times 101$ lattice.  
We stress that for a given configuration, this procedure
determines {\rm exactly} the number of sites that the monomer
can visit, so that the distribution in Fig.~\ref{fig:treesizes} is
exact, up to statistical errors.   The average monomer tree size 
is $8.16\pm 0.01$ sites, and the distribution decays exponentially for 
large tree sizes, as $\exp (-0.064 \,s)$ (where $s$ is the tree size).
This implies that in a lattice of infinite size, a single monomer 
is always localized.  At low but finite monomer density, a collective 
mechanism involving more than one monomers is therefore 
needed for diffusion of dimers at large length and time scales.
This result should be contrasted with a recent similar analysis 
for the dimer model on the square lattice, which found that single 
monomers are only weakly localized, having a power law distribution 
with a diverging expectation value for the number of accessible 
sites~\cite{Square.Dimer.Monomer}. 

\begin{figure}[tb]
\epsfig{figure=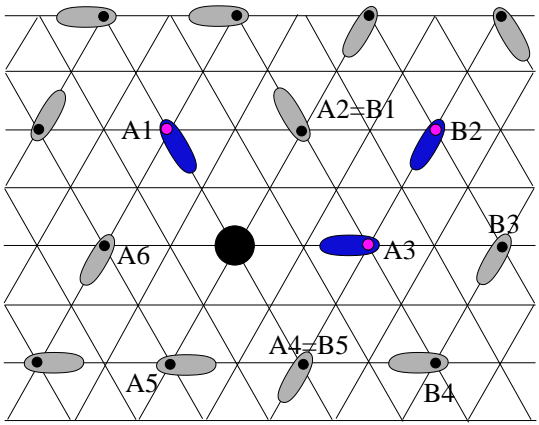,width=2.8in}
\caption{(Color online) Estimate of monomer tree size.  We only consider 
the dimers on sites in the same sub-lattice as the monomer.  
The dimers drawn are smaller than their actual size
in this picture.  The blue dimers are movable.  
The pink dots denote the sites reachable by the monomer
(i.e. the monomer tree).}
\label{fig:tree.heuristic}
\end{figure}

The exponential localization and average monomer tree size
on the triangular lattice can be understood by the following 
heuristic argument.  
For a configuration with a single monomer, we look at all the sites 
in the same sub-lattice as the monomer.
For the six next-nearest-neighbor sites 
in the same sub-lattice (sites $A_1$ to $A_6$ in 
Fig.~\ref{fig:tree.heuristic}), we assume that the dimers
on those sites each have an independent probability $1/6$ of
pointing in any of the six possible directions. This assumption 
should be fairly good, given the extremely short correlation length 
of dimer orientations in a fully packed triangular 
lattice~\cite{pocket-algor:Krauth}.   For each of those sites, 
if the monomer can reach that site, there are five more sites further 
out (for example, in Fig.~\ref{fig:tree.heuristic}, 
the site $A_3$ has neighbors $B_1$ through $B_5$),
each of which we assume has an independent $1/6$ probability
of being reachable~\footnote{The probabilities are not truly independent, 
both because there are short-range orientational correlations, 
and because the different branches overlap (e.g. $A_2=B_1$). 
We emphasize that this is only a rough estimate.}.
Continuing outwards, if we treat the different branches and orientation 
probabilities as independent, we have site percolation on a Cayley tree, 
with coordination number $z=6$   
and site occupation probability $p=1/6$. In this Cayley tree, the average
tree size is $7$, which agrees surprisingly well with our 
numerical result of $8.16\pm0.01$.  
Since cluster sizes for site percolation below the critical
point on the Cayley tree are exponentially distributed,
our heuristic argument also correctly predicts that 
large monomer trees are exponentially suppressed. 


\section{Localization of Monomer Clusters}
\label{Sec:localization} 
Let us define a monomer cluster to be localized, or confined, 
if the monomers can only reach a finite number of 
sites of the system, and delocalized, or deconfined, if some of them 
can reach an infinite number of sites. Our analysis in the preceding section
already shows that single monomers are always localized. 
In this section, we shall study clusters of monomers. 

\subsection{Localization with two monomers} 
\begin{figure}[tb]
\epsfig{figure=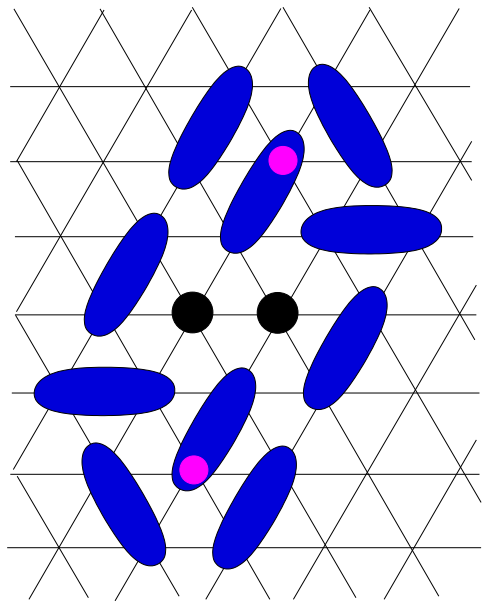,width=2.5in}
\caption{(Color online) A configuration in which two adjacent monomers are
confined. The small, light (pink) dots denote sites that can
be reached by the monomers.}
\label{fig:TrappedMMPair}
\end{figure}

A pair of nearest neighbor monomers can sometimes 
be localized. For example, in Fig.~\ref{fig:TrappedMMPair}, 
we show a configuration in which the two monomers can reach 
only a finite number of sites. It is clear that these two monomers 
sit on a swap cluster (defined in Sec.~\ref{Sec:intro}) that contains only 
two sites. Hence there is no swap move available. Furthermore, 
it is easy to check that the monomer trees of two monomers 
touch each other only at the current position 
of monomers. Each monomer is thus necessarily localized on their
individual tree. 

Our numerical simulations indicate that localization of a monomer pair 
only happens in roughly one-fourth of the configurations. 
In contrast to the single-monomer case, however, there is no
simple algorithm for determining if a pair of monomers is
localized, because the move of one monomer can change the
monomer tree of the other, as well as the configuration of swap
clusters. Very often, the two monomers appear to be trapped in a region 
for a long period of time, but after even longer times, the pair 
finds a way to break out of the region. Additionally, it is also 
possible that monomer pairs that appear delocalized for a given 
system size, would in fact be localized, if we considered the system
as a subset of an even larger lattice.  

We keep track of how many sites have had their occupation 
changed by a certain time---i.e. the dimer that initially covered 
the site moved at least once \footnote{even though the site
may be covered by the same dimer in the same way both in 
the initial state and in the final state}.  
We say that a configuration of monomers ``appears''
localized in a region of size $s$ at time $t$ if the number of sites 
whose occupation has changed is less than or equal to $s$. 
By creating many random configurations, each with one 
removed dimer, and running each up to time $t$,
we obtain the probability $p_s(t)$ that a given configuration 
with a monomer pair is localized within size $s$ at time
$t$, for any $s$.  By definition, for a given $t$, $p_s(t)$ is a 
monotonically increasing function of $s$ with 
$\lim_{s\rightarrow \infty}p_s(t) = 1$.  
In the limit $t\to\infty$, $p_s(t)$ approaches 
the probability  that a configuration is truly localized within size $s$. 

\begin{figure}[tb]
\epsfig{figure=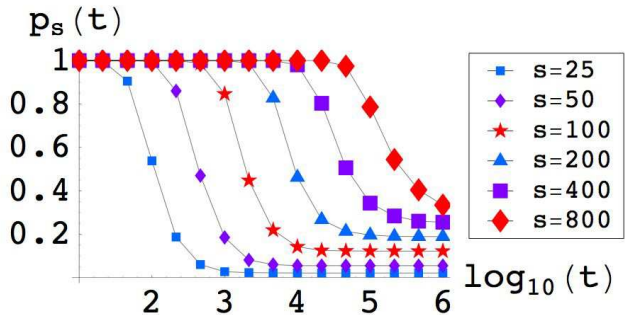,width=3.4in}
\caption{(Color online) Fraction of two-monomer configurations that ``appear'' to be 
confined within regions of size $s$ as a function of time.
The curves are for increasing $s$ from left to right.
The system size is $L=100$.}
\label{fig:confinement.n2}
\end{figure}

In Fig.~\ref{fig:confinement.n2} we show, for system of size $L=100$ 
with one monomer pair, $p_s(t)$ as a function of $t$, for different
values of $s$.  It is clear that each curve asymptotes to a nonzero value in
the infinite time limit. 
This indicates that at least a finite fraction of monomer pairs are confined. 
The asymptotic value $p_s(t \rightarrow \infty)$ increases appreciably 
with $s$, showing that monomer pairs have a broad range of localization sizes.

\begin{figure}[tb]
\epsfig{figure=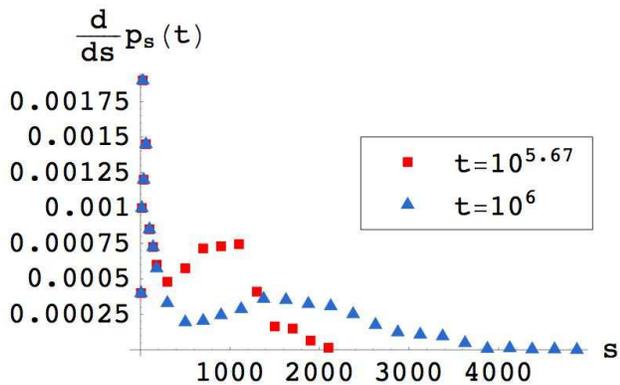,width=3.4in}
\caption{(Color online) The probability that exactly $s$ sites have moved at 
a given time $t$, for system size $L=100$, and two monomers. 
This is $\frac{d}{ds}p_s(t)$, where $p_s(t)$ is shown 
in Fig.~\ref{fig:confinement.n2}.
The two curves are for the times $t=10^{5.67}$ and $t=10^6$, and
lie on top of each other for $s<200$.}
\label{fig:moved.distrib.n2}
\end{figure}

The derivative of $p_s(t)$ with respect to $s$, $\frac{d}{ds}p_s(t)$, 
by definition, is the probability that at time $t$, the number of sites 
moved is exactly $s$.  In Fig.~\ref{fig:moved.distrib.n2} this probability
 is plotted as a function of $s$, at two different times. 
The red curve is at $t=10^{5.67}$, and the blue curve is at $t=10^{6}$, 
again for a system of size $L = 100$.   
It is clear that both curves contain two well separated peaks, 
a sharp and narrow peak at small $s$, and a secondary, wide peak 
at a larger value of $s$.  The narrow peak at small $s$ 
does not change with time, implying that the corresponding 
configurations are indeed localized. By contrast, the wider secondary
peak moves outwards with time, suggesting that 
the corresponding configurations are actually delocalized.

While Fig.~\ref{fig:moved.distrib.n2}  
suggests that there are delocalized states, it is difficult 
to precisely numerically determine the fraction of configurations 
that are delocalized. This is because, as already stated, 
localization is defined in the limit of infinite system 
sizes and infinite times. Any numerical definition 
of localization, however, needs to impose arbitrary cutoffs 
and criteria.  For a $240\times 240$ system, we simulated the 
system for $t=10^6$, and numerically defined a state 
as localized if both the monomers reached no new sites from 
time $t=10^5$ to $t=10^6$, and less than half of all sites 
had their occupation change.
With this numerical definition, we found 
that $21.5\pm1.3\%$ of the states are localized. This fraction 
varied as we changed the cutoffs and system size, ranging 
from 20\% to 25\%.  

\begin{figure}[tb]
\epsfig{figure=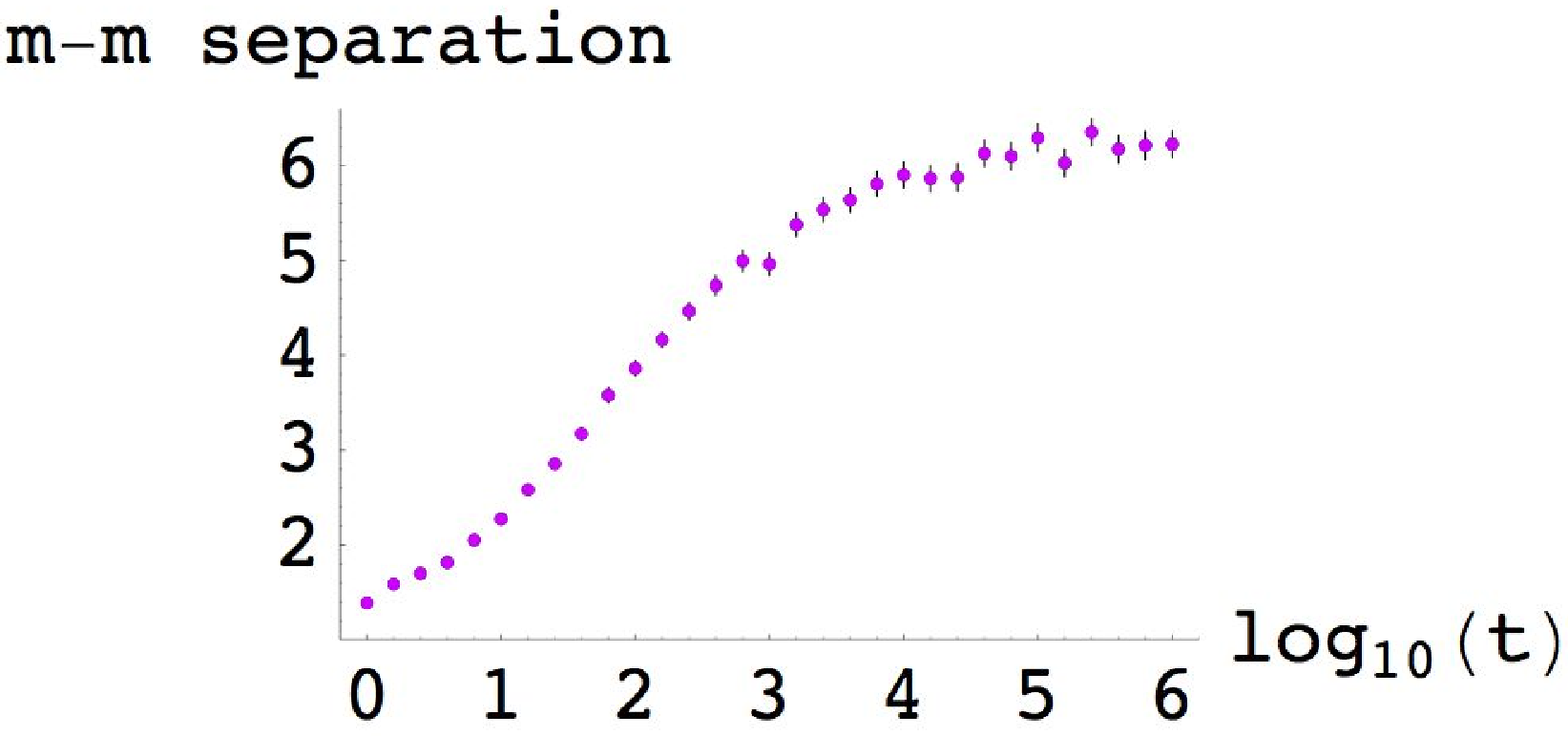,width=3.4in}
\caption{Average separation of a monomer-monomer pair in 
an otherwise fully packed lattice with $L=240$. 
This result is insensitive to system size.  }
\label{fig:monomono.separation.oneL}
\end{figure}

Since monomer trees are always of finite size, two monomers in a pair
(nearest neighbors in the initial state) cannot be widely separated.
In equilibrium, their separation should be proportional to the linear size 
of monomer trees. In Fig.~\ref{fig:monomono.separation.oneL}, we see the
average monomer-monomer separation as a function of time
for a system with two monomers. While the horizontal axis
spans six decades of time, the average monomer-monomer separation 
asymptotes at roughly 6 lattice spacings by $t=10^4$.
This verifies that monomer pairs are indeed bound. 
The asymptotic value of 6 lattice spacings 
is consistent with the average monomer tree size, and
does not change as we vary the system size. 
This result confirms the mechanism presented in Sec.~\ref{Sec:intro}: 
A pair of monomers can diffuse only through a two-monomer 
collective moves, which involves both swap and glide moves.

While we have not rigorously proven that monomer-pair localization 
actually happens on infinite lattices with finite probability, our numerical 
results presented above and in later sections strongly suggest that 
it is the case.   On the other hand, it turns out that the dimer system 
with only one dimer pair is never ergodic with the constraint of no rotation:    
There are three orientations of dimers: horizontal, northeastern, and 
northwestern. The orientation of each dimer is conserved by
the dynamics, so the number of dimers with each orientation does not 
change over time.  

Furthermore, it turns out that the system is not even ergodic 
within a sector with numbers of dimers of each orientation fixed, 
Recall that the system with $L$ even can be divided into
four sub-lattices---see Fig.~\ref{fig:sublattice}---and that
glide moves leave a monomer in the same sub-lattice.
Suppose, without loss of generality, that the removed dimer is 
horizontal, with one monomer in sub-lattice A, and the other in sub-lattice B. 
Subsequent glide moves will leave the monomers in their sub-lattices. 
Therefore, wherever the monomers
reconnect to one other, they will still form a missing dimer
of horizontal orientation. Hence the only possible swap move is to 
move a horizontal dimer, putting the monomers in sub-lattices 
C and D respectively (again, see Fig.~\ref{fig:sublattice}). 
Likewise, monomers in sub-lattices C and D can also only 
form missing horizontal dimers. Therefore the dynamics 
never allow a swap move of a non-horizontal dimer---such 
dimers can only undergo glide moves. It then follows that 
along every northeast (northwest) lattice line, the number 
of northeast-oriented (northwest-oriented) dimers is conserved. 
For a system with linear size $L$, this means a total of $2 L$ 
conserved quantities.  


\subsection{Localization with three or more monomers}

\begin{figure}[tb]
\epsfig{figure=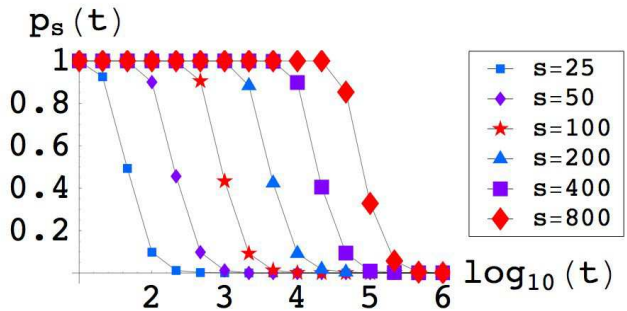,width=3.4in}
\caption{(Color online) Fraction of three-monomer configurations 
that ``appear'' 
to be confined, as a function of time, in regions of size $s$.
The curves are for increasing $s$ from left to right.
The system size is $L=100$.}
\label{fig:confinement.n3}
\end{figure}

\begin{figure}[tb]
\epsfig{figure=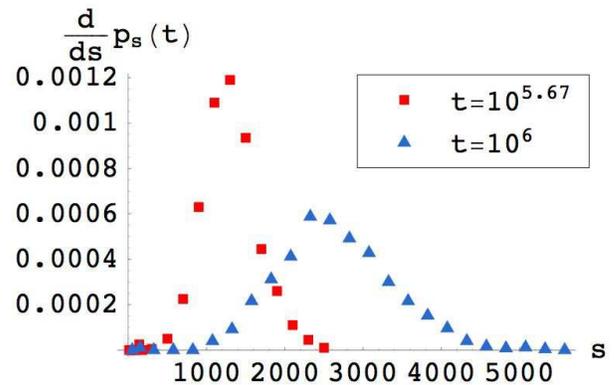,width=3.4in}
\caption{(Color online) 
The probability that exactly $s$ sites have moved at a given time $t$, 
for system size $L=100$, and three monomers. 
This is $\frac{d}{ds}p_s(t)$, where $p_s(t)$ is shown 
in Fig.~\ref{fig:confinement.n3}.
The two curves are for the times 
$t=10^{5.67}$ and $t=10^6$.}
\label{fig:moved.distrib.n3}
\end{figure}

The analogues of Figs.~\ref{fig:confinement.n2} 
and~\ref{fig:moved.distrib.n2} for the three monomer case
are shown in Figs.~\ref{fig:confinement.n3}
and~\ref{fig:moved.distrib.n3}.
In Fig.~\ref{fig:confinement.n3}, we see that the
percentage of states confined in a region of size $s$
appears to go to zero as $t\to\infty$ for any $s$.
And in Fig.~\ref{fig:moved.distrib.n3}, there is no
noticeable localized peak. So the numerical simulations seem to
indicate that all states with three connected monomers are
delocalized. 

\begin{figure}[tb]
\epsfig{figure=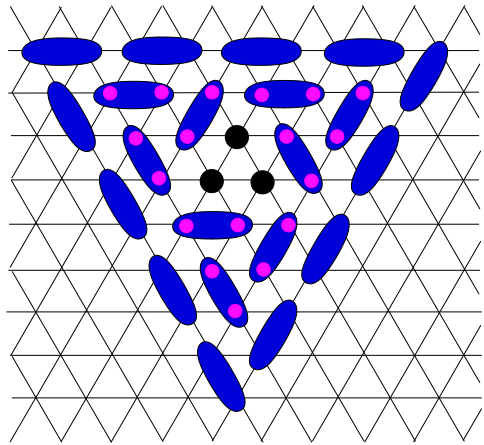,width=2.5in}
\caption{(Color online) A configuration in which three connected monomers are
confined.  The small, light (pink) dots denote sites that can
be reached by the monomers.}
\label{fig:TrappedMMMTri}
\end{figure}

\begin{figure}[tb]
\epsfig{figure=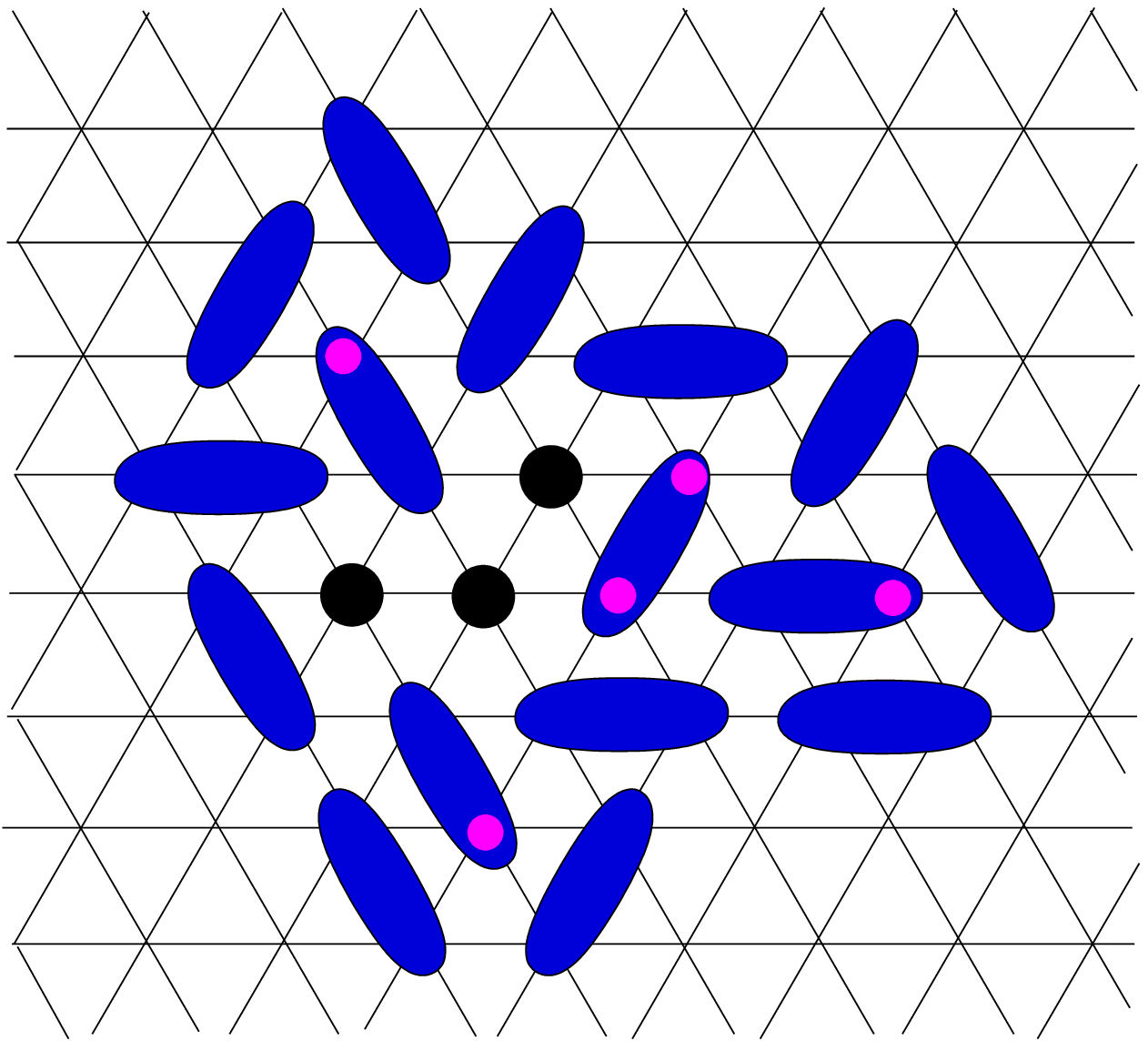,width=2.5in}
\caption{(Color online) A configuration in which three connected monomers are
confined.  The small, light (pink) dots denote sites that can
be reached by the monomers.}
\label{fig:TrappedMMMBent}
\end{figure}

\begin{figure}[tb]
\epsfig{figure=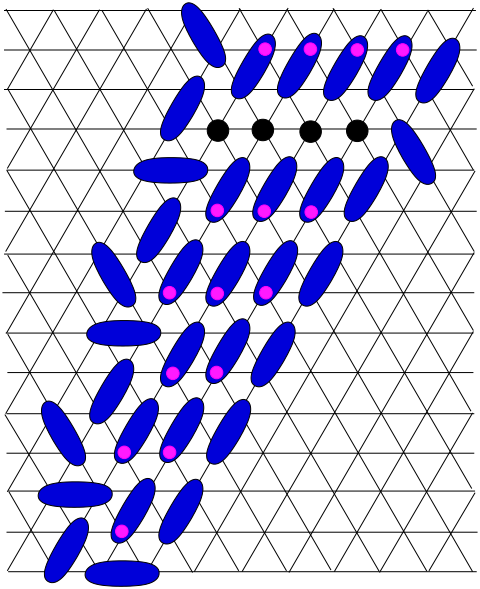,width=2.5in}
\caption{(Color online) Bottom half of a 
configuration in which four connected monomers in a
line are confined. The omitted top half is identical to
the bottom half below the four monomers, but
rotated $180^o$.  The small, light (pink) dots denote sites that can
be reached by the monomers.}
\label{fig:TrappedM4Line}
\end{figure}

It is actually possible, however, for clusters with arbitrary numbers of 
monomers to be localized. Two such examples for 3-monomer clusters 
are shown in Figs.~\ref{fig:TrappedMMMTri} and~\ref{fig:TrappedMMMBent}. 
Fig.~\ref{fig:TrappedM4Line} shows the bottom half of a configuration 
in which a 4-monomer cluster is confined (the undisplayed top half is 
identical to the bottom half, up to a rotation of $180^{\circ}$). 
The latter configuration can be generalized in a straightforward way to
arbitrary numbers of monomers in a straight line. In equilibrium, however,
localized configurations with three or more connected 
monomers appear with extremely small (nevertheless remains 
finite for $L\rightarrow \infty$) probability, 
and are actually never seen in our simulations. 
Visual inspection of the dimer dynamics confirms that configurations with 
a 3-monomer cluster are always delocalized in practice.

\section{Anomalous Diffusion of Monomers}

\label{Sec:diffusion}

Our analysis in the preceding section shows that a single 
monomer is always localized, while two or three-monomer clusters 
have finite probability to be delocalized. In this section, 
we shall characterize the diffusion of these delocalized 
monomer clusters.

\subsection{Diffusion of monomer clusters}

Our analysis in the preceding section shows that about 
20\%-25\% of two monomer pairs are localized. We are interested 
in the diffusion behavior of the delocalized monomer pairs.
Unfortunately in numerical simulations, it is difficult 
to reliably distinguish delocalized cases from localized 
ones. To avoid this difficulty and obtain better numerical 
results, we primarily simulated diffusion of three monomer
clusters.

\begin{figure}[tb]
\epsfig{figure=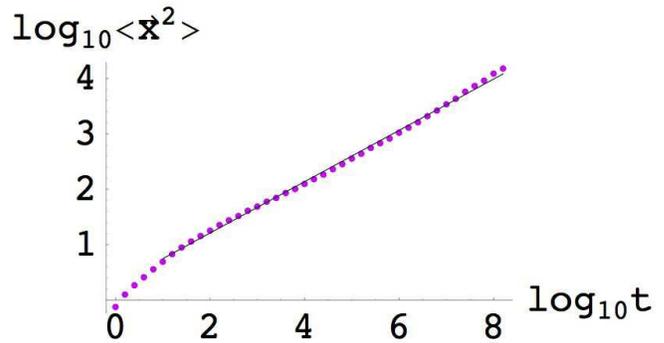,width=3.4in}
\caption{Monomer diffusion at large times. ($L=501$, 3
monomers.)}
\label{fig:longrun.monomer.r2.L501}
\end{figure}

We consider the diffusion of monomers with an initial
$3$-monomer cluster in a system of size $L=501$. The result 
is shown in Fig.~\ref{fig:longrun.monomer.r2.L501}. We 
averaged over $1500$ independent samples, running each 
sample for time $t=10^{8.2}$. This simulation took
seven days on a computer with a 2.0 GHz processor. 
We observed anomalously slow diffusion of monomers, with 
the average displacement square $\langle \vec{x}^2 \rangle$ scaling as 
\begin{equation}
\label{eq:r2diff.mono}
\langle \vec{x}^2 \rangle \propto t^\beta\, , \quad
\beta=0.46\pm0.06. 
\end{equation}

\noindent The data in Fig.~\ref{fig:longrun.monomer.r2.L501} 
shows some deviation from pure power law behavior, but the behavior 
is clearly subdiffusive over seven decades of time. 
The error bar in $\beta$ is
obtained from the variations in the slope
over different time ranges.
Simulations with four monomers 
give a similar curve, with a similar exponent
($\beta=0.44\pm0.04$), as do shorter simulations
with two monomers.

The initial configurations are prepared by starting with an
equilibrium state that is either fully packed, or 
contains only one vacancy, and then removing random adjacent 
dimers to form a connected hole.
While the fully packed configuration is in equilibrium, however, 
the configuration generated by the removals is not, as we
can see by looking at the average monomer tree size. 
The monomer tree construction is
most useful for a configuration with a single
monomer, but we can still define monomer trees for a configuration
with multiple monomers, by determining for each monomer 
the set of sites reachable by glide moves, while holding
all other monomers fixed. 
These trees give a rough characterization of the space
available to each monomer, but no longer tell us which sites
a monomer can ultimately reach, both
because they neglect swap moves, and because
each monomer's moves may change the trees of the other monomers.

\begin{figure}[tb]
\epsfig{figure=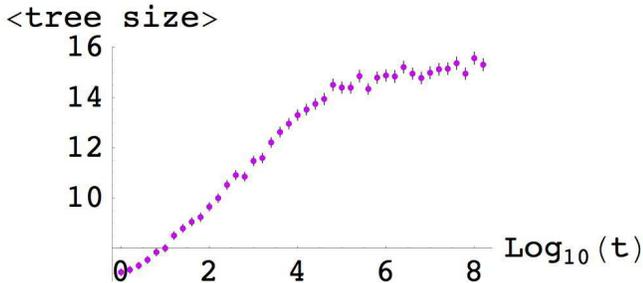,width=3.4in}
\caption{Growth of the average monomer tree size at large times. 
($L=501$, 3 monomers.)}
\label{fig:longrun.treesize.L501}
\end{figure}

In Fig.~\ref{fig:longrun.treesize.L501}, we see that for a configuration with three monomers, the average tree size grows very slowly (logarithmically) with time, before asymptoting to a roughly constant value.  This shows that it takes some time for our initial configuration to reach an equilibrium. The average monomer tree sizes grow in
a similar fashion for configurations with 2 or 4 monomers. 


To understand how monomer clusters can diffuse at long
length scales, let us first consider a configuration with two 
monomers that are nearest neighbors, such as the configurations 
shown in Figs.~\ref{fig:swap.picture.treestouch} 
and~\ref{fig:swap.picture.treesapart}. 
Numerical simulations show that the individual monomer trees
are exponentially distributed, even for configurations with
multiple monomers. Hence, 
{\em glide moves alone are not sufficient to allow a monomer 
pair to diffuse}. 

On the other hand, if we prohibit single glide moves and only 
allow swap moves 
and double-glide moves (including the latter since they
leave the monomers connected), then monomer pairs are
always confined to their swap cluster (defined in Sec.~\ref{Sec:intro}). 
Numerical simulations show that the swap cluster is also exponentially 
distributed, with an average of $17.89\pm0.02$ 
sites---the distribution is shown in Fig.~\ref{fig:swap.distribution}. 
Therefore, {\em swap moves alone are also not sufficient to 
allow a monomer pair to diffuse}. 
{\em Hence both glide and swap moves are essential 
to large-scale diffusion of monomer clusters.}

For a two-monomer configuration, such as shown in 
Figs.~\ref{fig:swap.picture.treestouch}
and~\ref{fig:swap.picture.treesapart}, the
two monomers can either glide separately 
along their individual monomer trees, or move together by 
swap moves. It is clear, however, that immediately after 
a glide move takes place, the 
two monomers stop neighboring each other and swap moves
are no longer possible. Subsequently, the two monomers perform
separate random 
walks on their own monomer trees, much like isolated 
monomers in an otherwise fully packed lattice. This situation remains 
true so long as two monomers do not become nearest neighbors, 
an event which we shall call ``reconnection of a monomer pair''.
For large monomer trees, the probability of a reconnection
at any given time is small.
After reconnecting, a swap
move necessarily changes the structure of the monomer trees.
The monomer can then perform random walks in their new monomer 
trees, until they eventually reconnect again. It is clear 
from this picture that for most of the time steps, the two monomers 
remain separated, performing random walks on their individual 
trees. A monomer pair has to overcome entropic barriers 
(log of monomer tree sizes) in order to reconnect and perform 
swap moves. This entropic barrier partially explains the 
slow diffusion seen in the simulations.

There are two possible ways that two monomers can reconnect: a) If 
the two monomer trees touch each other only in one place,
as in Fig.~\ref{fig:swap.picture.treesapart} then 
the only way for two monomers to reconnect is for each 
of them to, at exactly the same time, go back to the original sites 
where they separated. After reconnecting, they then with finite probability
perform swap moves, changing the monomer trees. The two 
monomers then may separate again and perform random walks 
in their new trees. 
b) If two monomer trees touch at more than one 
place (multiple junctions), as shown in 
Fig.~\ref{fig:swap.picture.treestouch},
then the two monomers may reconnect 
if they arrive at any of these places simultaneously. 
Here the most interesting possibility is that two monomers 
may reconnect inside a swap cluster different 
from the one they started with. This possibility provides 
a mechanism for a monomer pair to ``tunnel'' between different 
swap clusters, not unlike a Cooper pair tunneling between 
neighboring superconducting grains.

A monomer pair would be able to diffuse at large 
length and time scales only if these tunneling events happen with 
sufficiently high probability.   To characterize this probability, 
we start from a random fully packed configuration,
remove a single dimer randomly, and numerically check the number 
of reconnection sites, i.e. number of contacts between two monomer trees. 
We find that $38.7\pm0.1\%$ of the time, the two monomer trees
touch only at the original locations of the 
monomers (as in Fig.~\ref{fig:swap.picture.treesapart}),
while $36.0\pm0.1\%$ of the time, there is another
location where the monomers can meet. 
However, the quantitative relation between these probabilities and 
the diffusion behavior of monomers is not easy to obtain. 

This ``swap-tunneling'' mechanism is responsible for $2$-monomer 
diffusion, but it is unclear whether this mechanism 
is also responsible for monomer diffusion in a system at 
low monomer density. To explore this issue, we have also 
simulated the diffusion of larger monomer clusters. Visual 
inspection of the diffusion dynamics shows that 
1) For a three-monomer cluster, most of the time two 
monomers remain relatively close to each other and are mutually 
connected by glide moves, while the third one is very 
often far away. Furthermore, a three-monomer cluster is 
always localized if no swap moves are allowed, or if glide 
moves are not allowed. Therefore it appears that the two 
monomer ``swap-tunneling'' dynamics dominates the diffusion 
of a three-monomer cluster. 2) Our numerical simulations 
clearly show that monomer clusters of six or fewer connected 
monomers are localized if swap moves are prohibited. This 
strongly suggest the importance of swap move in large-scale 
diffusion of monomer clusters. 3) Four-monomer clusters 
can be delocalized even if glide moves are disallowed,
showing that swap moves alone provide
a mechanism for monomer diffusion at low monomer density. 
However, visual inspection of the simulations shows that when 
both swap and glide moves are allowed, larger clusters 
of (four or more) monomers tend to separate into smaller 
clusters containing one or two monomers. This separation 
is entropically favorable: there are many more possible 
glide moves than possible swap moves. More importantly, 
it shows that the ``swap-tunneling'' mechanism of two-monomer 
clusters is indeed the most important mechanism for the 
large-scale transport of monomers at high packing densities. 
This conclusion is also supported by our study of dimer 
diffusion at low by finite monomer density, discussed in 
Sec.~\ref{Sec:dimer-diffusion}.


\subsection{Reconnection times}

\begin{figure}[tb]
\epsfig{figure=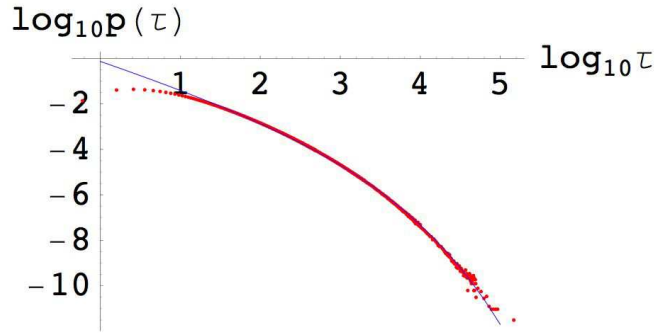,width=3.4in}
\caption{The distribution of reconnection times,
fitted to a stretched exponential,
$\tau^{-\alpha} \exp (-\beta \tau^{1/3} )$, 
with $\alpha=1.13\pm 0.04$, 
and $\beta=0.30\pm 0.02$.}
\label{fig:reconnection.distribution.fit}
\end{figure}

The argument of the preceding subsection 
indicates that in order for a monomer pair to diffuse, it 
is essential for the two monomers to reconnect 
in a different swap cluster. We now study the distribution 
of time separations between successive reconnection events 
for pairs of monomers in more detail. Let us first define 
precisely reconnection events for configurations with only two monomers. 
Suppose that at time $t_i$, the two monomers are on neighboring sites. 
We identify their swap cluster, and then 
allow the system to evolve.  We say that a reconnection event 
happens at time $t_{i+1}$ if at this time step the two monomers lie on
neighboring sites {\em in a different swap cluster}
than at time $t_i$. We define 
the time difference $\tau_{i}\equiv t_{i+1}-t_i$ as the reconnection 
time. We then recalculate the new swap cluster of the monomer 
pair, and repeat the process. To make sure that we study 
the equilibrium properties of the system with a monomer pair, 
we simulate the system for a long amount of time ($t_{\rm 
init}$), before collecting 
a sequence of reconnection times
for a smaller time window ($t_{\rm coll}$).
The distribution of reconnection 
times thus obtained is shown in 
Fig.~\ref{fig:reconnection.distribution.fit}. 

There is a simple way to understand this reconnection time 
distribution: When we have two connected monomers, each 
will have 
its own monomer tree, with sizes $\ell_1$ and $\ell_2$ respectively. 
Let us assume that these two trees 
are identically and independently distributed, each with 
the probability distribution
\begin{equation}
\label{eq:single.tree.distribution}
p(\ell ) = \ell^{-a} e^{-\ell /b}. 
\end{equation}
The above parameters $a$ and $b$ shall be determined by
fitting the curve and may be slightly different than 
those for the isolated monomer case. We further assume that 
after two monomers separate, each of them quickly independently 
equilibrates in its own tree, and that the two trees touch 
at $m$ points in other swap clusters,
where $m$ is some fixed number (i.e. does 
not scale with $\ell_1$ and $\ell_2$).

Under these assumptions, at any given time, the probability 
that the two monomers are adjacent is $m/(\ell_1\ell_2)$. 
The probability that they reconnect for the first time at 
time $\tau$ (an integer) is thus 
\begin{equation}
\left( 1 - \frac{m}{\ell_1\ell_2} \right)^{\tau-1}
\frac{m}{\ell_1\ell_2}
\approx
\frac{m}{\ell_1\ell_2} e^{-\frac{m\tau}{\ell_1\ell_2}}. 
\end{equation}

\noindent Averaging over the distributions of $\ell_1$ and 
$\ell_2$, we get the probability of first reconnecting at 
time $\tau$ to be
\begin{equation}
\int d\ell_1 \int d\ell_2 \ p(\ell_1) p(\ell_2)
\frac{m}{\ell_1 \ell_2} e^{-\frac{m \tau}{\ell_1\ell_2}}. 
\label{integral-11}
\end{equation}

\noindent Note that this is, by definition, the distribution of 
reconnection times for a monomer pair. 

Performing saddle point approximations for both 
the $\ell_1$ and $\ell_2$ integrals in Eq.~(\ref{integral-11}), 
we find that the distribution 
of reconnection times is proportional to
\begin{equation}
\tau^{-(2a+1)/3} \exp\left(-\frac{3(m\tau)^{1/3}}{b^{2/3}}\right). 
\end{equation}

In Fig.~\ref{fig:reconnection.distribution.fit}, we see
that the distribution is indeed fit well with such a stretched 
exponential, $\tau^{-\alpha} \exp(-\beta \tau^{1/3})$, with 
$\alpha=1.13\pm 0.04$, 
and $\beta=0.30\pm 0.02$. If we fit the single-monomer distribution 
in Fig.~\ref{fig:treesizes} with 
Eq.~\ref{eq:single.tree.distribution},
we get $a=1.4$ and $b=13$; if we then assume $m=1$, we get 
$\alpha=1.27$, and $\beta=0.54$, which matches the fitted 
values reasonably well, given the approximations made.

We may try to understand the anomalous diffusion of monomer
pairs in terms of sporadic swap moves and tunneling, separated by glide 
moves that do not contribute to large-scale diffusion
(separating out the glide moves that make up the 
tunneling events from those that do not).
The reconnection time therefore behaves much 
like the waiting time for a particle diffusing in a random 
potential landscape with traps at each site, which separate 
succeeding hops. It is well known that a waiting time distribution 
with a diverging average naturally leads to anomalous 
diffusion~\cite{review-anomalous-diffusion}. In our case, 
however, the average of the reconnection-time 
distribution (a stretched exponential) 
is clearly finite. We therefore conclude that the distribution 
of reconnection times that we see in our simulations does
not qualitatively 
explain the anomalous diffusion of monomers. 

\begin{figure}[tb]
\epsfig{figure=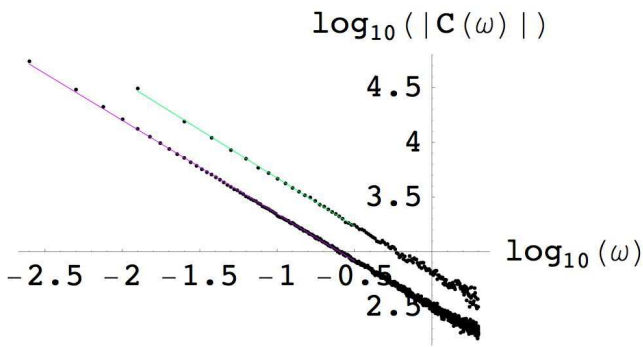,width=3.4in}
\caption{Correlations between reconnection times, in
frequency space. The upper (green)
curve is for $\{t_{\rm init},t_{\rm coll}\}=\{7.5 \times10^4, 2.5 \times 10^4\}$,
while the lower (purple) curve is for
$\{t_{\rm init},t_{\rm coll}\}=\{7.5 \times 10^5, 2.5 \times 10^5\}$.
$t_{\rm init}$ is the time used to equilibrate, and 
$t_{\rm coll}$ is the time window for measurement.
Both curves are for $L=250$.}
\label{fig:reconnection.correl.freq}
\end{figure}

We have also analyzed the correlation function of the reconnection time 
sequences.   For a given sequence $\{\tau_1,\tau_2,\tau_3,\ldots\}$
of reconnection times,
the correlation function is defined to be
\begin{eqnarray}
C(j) & \equiv & \langle \tau_{i+j} \tau_i \rangle - 
\langle \tau \rangle^2, \\
C(\omega) & = & {\frac{1}{\sqrt{M}}} \sum_{j=1}^M 
e^{i\omega (j - 1)} C(j) \quad ,
\end{eqnarray}
where $M$ is the maximum value of $j$, 
the correlation distance. The correlation function in 
frequency space, as shown in Fig.~\ref{fig:reconnection.correl.freq}
for two different collection time windows (values of $\{t_{\rm 
init},t_{\rm coll}\}$), depends on the frequency as a power law. 
While the prefactor of the correlation function depends on the time window, 
the slope does not. In the frequency space, the correlation function scales as
\begin{equation}
C(\omega) \propto \omega^{-(1-\gamma)} ,\quad
\gamma=0.14\pm0.01. 
\end{equation}
This long-range correlation in reconnection times should be related to 
the anomalous diffusion behavior of monomers. In particular, it may be
related to the probability of a monomer pair revisiting its
initial swap cluster. 
A quantitative understanding of this correlation, however, is still lacking. 

\section{Dimer diffusion at finite monomer densities}
\label{Sec:dimer-diffusion}

Our study of monomer diffusion suggests that coordinated 
``swap-tunneling'' motion of 
monomer pairs constitutes the basic mechanism for diffusion of monomer 
clusters.  In this section, we study diffusion of dimers at finite but 
low monomer density and show that it can also be understood 
in terms of monomer pairs. 

The first question we need to address is, for a given monomer 
density, what is the density of monomer pairs? Let us define 
that two monomers form a pair if they, with all other monomers 
fixed, can be made nearest neighbors by glide move of 
dimers. Clearly this is possible if and only if two monomer 
trees touch each other at one or more sites, as illustrated 
in Figs.~\ref{fig:swap.picture.treestouch} 
and~\ref{fig:swap.picture.treesapart}.

A rough estimate of the probability that a second monomer 
touches a given monomer tree can be obtained as follows. 
We first want to count the number of distinct neighbors 
of the sites in the monomer tree. The site that the monomer 
begins at has $6$ neighbors. Each new site in the tree 
adds $5$ new neighbors (as one neighbor
along the edge of the tree has already been counted).
The new neighbors may not all be distinct, since sites of 
the tree may have overlapping neighbors not on the edges 
of the tree. Ignoring such overlapping cases, and using 
the fact that the average size of monomer trees is $8.16$, 
we get we have $6+5(7.16)\approx 42$ neighbors. If we further 
assume that monomer positions are independent and 
uncorrelated~\footnote{In Ref.~\cite{dimer-triangular-Fendley} 
it was found that monomer--monomer 
correlations are extremely short-ranged, with a correlation 
length less than one lattice step.}, then the probability 
that the given monomer forms a pair with some other monomer 
is $1-(1-\rho_m)^{42}$. 

\begin{figure}[tb]
\epsfig{figure=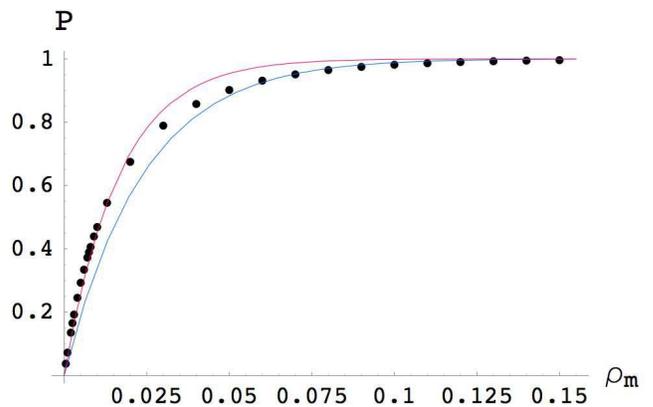,width=3.4in}
\caption{The probability that a given monomer is in a pair with 
some other monomer, as a function of the monomer density. 
The lower curve is $1-(1-\rho_m)^{42}$, and the upper curve is
$1-(1-\rho_m)^{60}$. }
\label{fig:monomer.pair.density}
\end{figure}

To test this simple estimate, we 
generate random configurations at finite monomer densities using 
the pivot algorithm, and count the total number of monomer 
pairs. As shown in Fig.~\ref{fig:monomer.pair.density}, 
the numerical results for the probability that a monomer 
is in a pair agree well with this formula (never differing 
by more than a factor of $2$, even at the lowest density
tested, $\rho_m=0.0005$). A better fit, $1-(1-\rho_m)^{60}$,
also shown in Fig.~\ref{fig:monomer.pair.density}, can 
be obtained by varying the effective number of tree neighbors. 

We find that at a monomer density of around $2\%$, the majority 
(about $70\%$) of monomers already form pairs. On the other 
hand, at much lower monomer densities, the probability that 
a given monomer forms a pair with some other monomer is 
linear in $\rho_m$. Therefore for $\rho_m \ll 0.02$, the 
monomer pair density scales as $\rho_m^2$, while for $\rho_m 
\gg 0.02$, the monomer pair density scales as $\rho_m$.

\begin{figure}[tb]
\epsfig{figure=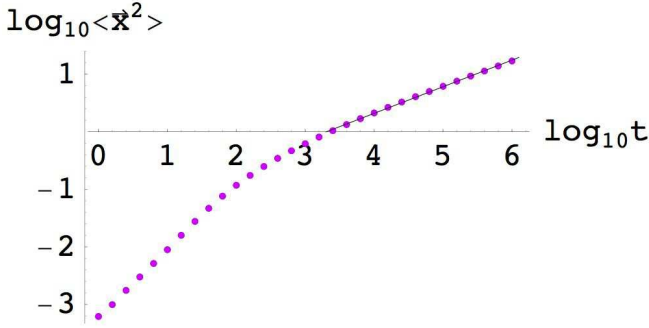,width=3.4in}
\caption{The average dimer displacement squared
scales as $\langle \vec{x}^2 \rangle = k\, t^{0.47}$, averaging 
over both configurations and dimers, for a $241 \times 241$ system 
with $\rho_m=0.004$.  } 
\label{fig:finitemonoden.dimerdiff.sample.rho004}
\end{figure}

To study dimer diffusion, we generate equilibrium configurations 
at a finite monomer density using the pivot algorithm. We 
then evolve the system, keeping track of the value of $\langle 
\vec{x}_i^2\rangle$ for each dimer, summing over all dimers (including 
those never moved) in the configuration, and averaging over 
different configurations. A representative plot of results 
for $\rho_m=0.004$ is shown in 
Fig.~\ref{fig:finitemonoden.dimerdiff.sample.rho004}. 
The short time behavior (for $t \leq 10^2$) is dominated 
by glide moves of monomers on their individual trees. 
At longer time scales (for $10^3 \leq t \leq 10^6$), we find the scaling 
\begin{eqnarray}
\langle \vec{x}^2 \rangle = k \, t^{\tilde{\beta}}, 
\hspace{3mm} 
\tilde{\beta} = 0.47\pm 0.02. 
\label{dimer-diffusion}
\end{eqnarray} 
We have also simulated monomer densities between 
$0.0005$ and $0.015$, and found that $\tilde{\beta}$ is roughly constant.
For larger $\rho_m$, $\tilde{\beta}$ increases with $\rho_m$,
reaching $0.9$ for $\rho_m=0.4$.

\begin{figure}[tb]
\epsfig{figure=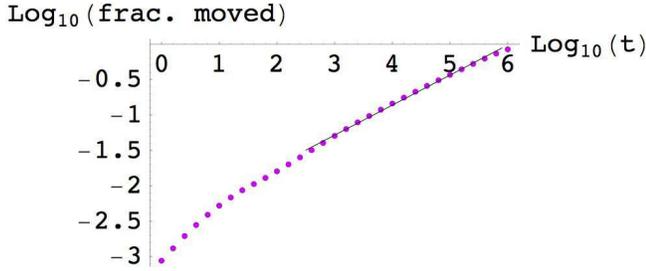,width=3.4in}
\caption{The fraction of dimers that have moved at least
once, as a function of time, for a $241 \times 241$ system
with $\rho_m=0.004$.}
\label{fig:finitemonoden.percentage.moved.rho004}
\end{figure}

\begin{figure}[htb!]
\epsfig{figure=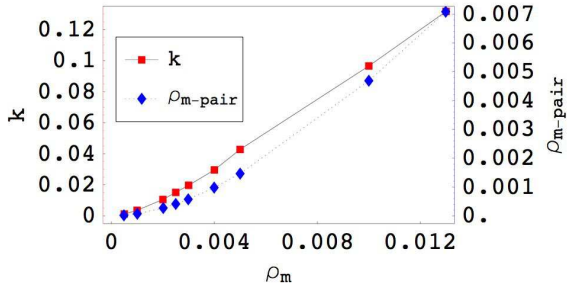,width=3.1in}
\caption{The coefficient $k$ in $\langle \vec{x}^2 \rangle = k \, t^{\tilde{\beta}}$
for diffusion of dimers (squares, left axis), and
the density of monomer pairs, $\rho_{m-pair}$ (diamonds,
right axis), each as a function of monomer density $\rho_m$.
The ratio between these two quantities is roughly constant, 
varying by a factor less than $3$, while $\rho_m$ changes by two orders
of magnitude. }
\label{fig:finitemonoden.threegraph.comparison}
\end{figure}

We note that this dimer diffusion exponent $\tilde{\beta} \approx 0.47$ 
measured at low $\rho_m$ {\em equals} the 
monomer diffusion exponent $\beta = 0.46\pm 0.06$ found in 
Eq.~\ref{eq:r2diff.mono} within numerical precision. 
This supports our physical picture that diffusion of monomer pairs 
is the dominating mechanism of large scale transport.  At sufficiently low 
monomer densities ($\rho_m \leq 0.02$, for example), the monomer 
pair density is extremely low. Within reasonable time scales, then, 
dimer pairs remain well separated and do not touch each other.  
Hence we can treat the diffusion of each monomer pair separately. 
As time evolves, monomer pairs diffuse around their original
positions.  The radius squared of the region that the monomer pair 
visits scale as $t^{\beta}$, according to our simulation of monomer 
diffusion (Eq.~\ref{eq:r2diff.mono}).  If this region is compact 
(correct for 2d diffusion problems), the area of the region 
visited by a monomer pair should scale with the same exponent.  
Note that only dimers inside this region have moved. By contrast, 
dimers outside these regions are frozen at this particular 
time scale. Hence the system consists of growing ``active'' 
regions that have been visited by the monomer pairs, surrounded 
by ``inactive'' background. Hence the number of dimers that 
have moved is the same as the total area of these active 
regions, which scales as the total number of active regions 
multiplied by $t^{\beta}$.
Consistent with this, we have verified that the number of 
dimers moved scales as $t^\beta$, for all monomer
densities---a representative plot is shown in
Fig.~\ref{fig:finitemonoden.percentage.moved.rho004}. 
If we further assume that dimers within each active region 
on average diffuse up to distances of order of one,
then this is also the 
scaling behavior of dimer diffusion $\langle \vec{x}^2 \rangle$, 
given by Eq.~(\ref{dimer-diffusion}), hence $\tilde{\beta}=\beta$. 

\begin{figure}[tb!]
\epsfig{figure=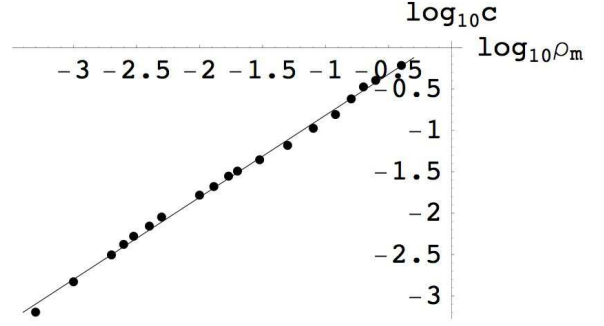,width=3in}
\caption{The coefficient $c$ in $\langle r \rangle = c \,
t^\delta$ for dimer diffusion, averaging over dimers and over
configurations, at long times, as a function of
monomer density. 
The straight line fit has a slope of $0.992\pm 0.013$.}
\label{fig:finitemonoden.dimerdiffr.coefficients}
\end{figure}

As a byproduct, this argument also predicts that the coefficient 
$k$ in Eq.~(\ref{dimer-diffusion}) should be linear in monomer 
pair density. In Fig.~\ref{fig:finitemonoden.threegraph.comparison} 
we plot both $k$, and the density of monomer pairs, $\rho_{m-pair}$, 
as functions of $\rho_m$, for low $\rho_m$. We see that both 
graphs are qualitatively similar. While both $k$ and $\rho_{m-pair}$ 
vary by a factor of $100$ as we vary $\rho_m$ from $0.0005$ 
to $0.013$, the ratio between the two varies by less than 
a factor of $3$. We thus conclude that the variation of 
the monomer pair density is primarily responsible for the 
variation of $k$, and that monomer pairs are indeed responsible 
for large-scale transport of dimers.

We have also looked at the averaged dimer displacement 
$\langle r \rangle = \langle |\vec{x}| \rangle$ (rather 
than the average of displacement squared) at finite monomer 
densities. We again focus on the behavior at larger 
times, and fit the results to 
\begin{eqnarray}
\langle r \rangle = c \, t^\delta. 
\label{r-scaling}
\end{eqnarray}
Surprisingly, we find that $\delta$ is very close to the monomer 
diffusion exponent $\beta$, varying in the range $0.450\pm0.025$, 
for all monomer densities studied in the range of $0.0005 <\rho_m < 0.40$. 
The equality of the exponent $\delta$ with the exponent $\beta$ of 
Eq.~\ref{eq:r2diff.mono} is expected for low $\rho_m$, 
by the same arguments we presented 
earlier. However, at higher monomer densities, we find no 
reason why the exponent $\delta$ in Eq.~(\ref{r-scaling}) 
should remain unchanged---at these densities, most monomers 
form pairs and the physical picture where the space consists 
of isolated active regions embedded in an inactive background, 
is no longer valid. Most dimers end up moving by the onset 
of anomalous diffusion of dimers. Probably even more puzzling 
is that the coefficient $c$ in Eq.(\ref{r-scaling}) is linear 
in $\rho_m$ over three decades in $\rho_m$, as shown in 
Fig.~\ref{fig:finitemonoden.dimerdiffr.coefficients}.
The best fit line on a log--log scale has a slope of $0.992\pm 
0.013$, indicating a linear dependence of $c$ on $\rho_m$. 
An understanding of this scaling behavior is lacking.



\section{Conclusion}
\label{Sec:conclusion}
In this work we have studied anomalous diffusion of monomers 
and dimers in the triangular lattice dimer model, subject to the constraints 
that dimers cannot rotate and that each site can only be occupied by one dimer.  
We have identified monomer pairs as the basic degree of freedom 
for large-scale transport of monomers and dimers, and have proposed 
a ``swap-tunneling'' mechanism that involves a subtle interplay between 
swap moves and glide moves.   A quantitative understanding of the anomalous 
exponent for monomer diffusion, however, remains elusive.  It will be interesting 
to further explore whether this intricate vacancy dynamics is relevant to vacancy 
diffusion in glassy systems as well as in densely packed granular aggregates.  
Finally we note that our model exhibits no equilibrium jamming transition at finite monomer density: As long as monomer density is finite, there is always probability 
one to find two-monomer clusters (as well as larger clusters) in an infinite system.  
According to the results of our work, then, dimers always diffuse anomalously 
as Eq.~(\ref{dimer-diffusion}), at time scales longer than $10^3$.   At even longer 
time scales, monomer clusters with large sizes come into effects and dimers may 
eventually diffuse normally.  

\vspace{-5mm}

\acknowledgments
This work was supported in part by grants ACS PRF 44689-G7 (XX) and
NSF DMR-0645373 (JMS).  MJB and XX thank R. Kenyon for interesting discussion. 
The authors acknowledge the hospitality of thank Aspen Center for Physics, where the collaboration began, 

\bibliography{reference-dimer-1}


\end{document}